\documentstyle[12pt]{article}

\newcommand{\be}{\begin{equation}}
\newcommand{\ee}{\end{equation}}
\newcommand{\bea}{\begin{eqnarray}}
\newcommand{\nn}{\nonumber}
\newcommand{\eea}{\end{eqnarray}}

\newcommand{\xd}{\chi}

\newcommand{\yy}{W}

\begin {document}

\begin{flushright}
NEIP-97-010\\
hep-th/9808098 \\
\end{flushright}

\begin{centering}
\bigskip
{\Large \bf 
T-duality for boundary-non-critical 
point-particle and string quantum mechanics}\footnote{To appear in the 
proceedings of {\it Quantum Aspects of Gauge Theories, 
Supersymmetry and Unification}, Neuch\^atel, 18-23 September 1997.}

\bigskip
\medskip

{\bf Giovanni AMELINO-CAMELIA}\\
\bigskip
Institut de Physique, Universit\'e de Neuch\^atel,\\
rue Breguet 1, CH-2000 Neuch\^atel, Switzerland\\
and\\
Theoretical Physics, University of Oxford,\\
1 Keble Rd., Oxford OX1 3NP, UK

\end{centering}
\vspace{1cm}
     
\begin{abstract}
It is observed that some structures recently uncovered
in the study of Calogero-Sutherland models and anyons 
are close analogs of well-known structures of boundary conformal 
field theory.
These examples of ``boundary conformal quantum mechanics'',
in spite of their apparent simplicity,
have a rather reach structure, 
including some sort of T-duality,
and could provide useful frameworks for testing 
general properties of boundary conformal theories.
Of particular interest are the duality properties 
of anyons and Calogero-Sutherland particles in presence 
of boundary-violations of conformal invariance; these are
here briefly analyzed leading to the
conjecture of a general interconnection between 
(deformed) boundary conformal quantum mechanics, T-type duality, and
(``exchange'' or ``exclusion'') exotic statistics.
These results on the point-particle quantum-mechanics side
are compared with recent results on the action of T-duality on 
open strings that satisfy conformal-invariance-violating
boundary conditions. Moreover, it is observed 
that some of the special properties of anyon and 
Calogero-Sutherland quantum mechanics are also enjoyed by 
the M(atrix) quantum mechanics which has recently
attracted considerable attention.
\end{abstract}

\vspace{2cm}

\section{Introduction}

Over the last few years a large number of studies 
has been devoted to anyons \cite{anyrefs}
and Calogero-Sutherland particles \cite{calo,suth}, 
with emphasis on the fact that 
the former provide the canonical laboratory for the study of 
anomalous {\it exchange} statistics \cite{anyrefs}
while the latter exhibit
anomalous {\it exclusion} statistics \cite{haldane}.
Interestingly, these point-particle non-relativistic 
quantum-mechanical systems 
enjoy scale invariance if (as customary) the domain of the relevant
Hamiltonians is specified by the requirement that the wave 
functions be regular everywhere.
Some recent studies have examined the implications of
a certain class of ``deformations'' of these systems,
in which one consistently introduces
violations of scale invariance via 
self-adjoint extensions prescribing that the wave functions 
have isolated and square-integrable singularities at 
the boundary of the fundamental domain (the points of configuration
space where the positions of two of the particles coincide).
As I shall observe in the following,
in spite of its apparent simplicity
this type of boundary deformation of a scale-invariant quantum
mechanics leads to a rather reach structure, 
including many of the properties of boundary deformations of 
more complicated boundary-conformal field theories.

These ``boundary-conformal quantum mechanical systems''
(which I shall qualify as ``boundary-non-critical 
quantum mechanical systems'' once deformed by
non-conformally-invariant boundary physics)
could be useful as simple settings in which to
test ideas concerning more complicated theories 
with nontrivial boundary physics.
I shall illustrate this by observing that in the
Calogero-Sutherland and anyon problems 
one can find dualities that share
some of the properties of dualities holding
in open-string theories. 
This analogy still holds when
non-conformally-invariant boundary physics
is introduced, in which case the relevant
duality properties on the open-string side
are the ones here briefly reviewed in Sec.~3
(and originally derived in Ref.~\cite{gacnick}).

\section{Boundary-non-critical quantum mechanics and T-duality}

The possibility of self-adjoint extensions mentioned in the Introduction
can be illustrated very simply by looking at the Hamiltonian
\begin{equation}
{\cal H} \equiv H_{any}^\nu (r) 
= - {1 \over r} \partial_{r} (r \partial_{r})
+ { \nu^2 \over r^2} 
~.
\label{eqbe}
\end{equation}
This Hamiltonian not only describes the s-wave relative motion of 
a 2-anyon system with exchange-statistics parameter $\nu$,
but its eigensolutions are also simply related 
to the ones of a corresponding Calogero-Sutherland model.
The 2-body (relative-motion) Calogero-Sutherland model 
with exclusion-statistics parameter $\beta$ 
is described in the ``Calogero limit'' ({\it i.e.}
the infinite-radius limit for the circle on
which Calogero-Sutherland particles are constrained)
by the Hamiltonian
\begin{equation}
H_{CaSu}^\beta (x) 
= - {d^2 \over dx^2} 
+ { \beta (\beta -1) \over x^2} ~,
\label{casumy}
\end{equation}
which is related to $H_{any}^\nu$ by 
\begin{equation}
H_{CaSu}^\beta (x) 
= x^{-1/2} \, H_{any}^{\beta+1/2}(x) \, x^{-1/2}
\label{anytocasu}
\end{equation}
It is therefore sufficient to discuss the properties of the
Hamiltonian ${\cal H}$
defined in (\ref{eqbe}) in order to obtain insight in both
the s-wave sector of the (2+1-dimensional) 
2-body anyon problem and the full (1+1-dimensional) 
2-body Calogero-Sutherland problem.
(Generalizations to the N-body problems with $N>2$ have been 
discussed elsewhere, see {\it e.g.} Refs.~\cite{pap6p9,mto},
but it will be sufficient to consider the 2-body problems for 
the illustrative purposes of the present discussion.)

A first observation is that the ${\cal H}$-eigenproblem
is scale-invariant if the domain of ${\cal H}$ only includes 
wave functions that are regular everywhere, which is the preferred
choice in most of the related literature.
However, it is well known \cite{bourdeaumanu2,gacbak} that ``meaningful''
(self-adjoint ${\cal H}$) ${\cal H}$-eigenproblems
are obtained also from the more general class of 
boundary conditions\footnote{Eq.~(\ref{bc1}) and some of the 
following equations do not naively apply to the special limit $\nu = 0$.
Although very simple limiting prescriptions can be introduced to make
these formulas hold even in the $\nu = 0$ limit, for the present paper 
it suffices to focus on $\nu \ne 0$. The interested reader can find 
useful insight in the study of the special
case $\nu = 0$ in Ref.~\cite{nuzero}.
Concerning the parameter $w$ the reader should notice that,
as in Ref.~\cite{gacbak}, in the following only $w \ge 0$
are considered.}
\begin{equation}
\left[r^{|\nu|} \psi({r}) - w \rho^{2 |\nu|}
{d\left(r^{|\nu|} \psi ({r})\right)
\over d (r^{2 |\nu|})}\right]_{r=0}=0 ~,
\label{bc1}
\end{equation}
which in general is scale-dependent. The scale-independent
limits of (\ref{bc1}) are obtained at the special
values $0,\infty$ of the dimensionless parameter $w$, 
which characterizes the self-adjoint extension once
the reference scale $\rho$ is assigned.\footnote{In order to emphasize
this $\rho \leftrightarrow w$ interdependence (which has sometimes 
been missinterpreted in the related literature) in the following I also
use the compact notation $w_\rho$.}
In particular, the popular case of wave functions regular everywhere
corresponds to the scale-invariant limit $w \! = \! 0$,
and it is conventional to consider the other elements 
of the one-parameter family of domains described by (\ref{bc1})
as ``deformations'' of the ordinary regular-wave-function case.

Interestingly, 
these deformations that I have until now discussed
as encoded in boundary conditions at  $r \! = \! 0$
allow for a ``dual'' description in which the wave functions 
are all along taken to be regular everywhere ({\it i.e.}
satisfy  (\ref{bc1}) for $w \! = \! 0$)
and the boundary deformation is introduced
via a $\delta(r)/r$ contact interaction 
({\it i.e.} a boundary interaction, since particles only collide
at the boundary of the domain)
with running coupling $g(\mu)$.
Evidence of the equivalence 
of these two dual descriptions has emerged in several
studies \cite{pap6p9,mto,gacbak}.
By looking at the first few terms
in an appropriate perturbative expansion it has been shown 
how $g(\mu)$ captures the physics
of $\rho$ and $w$ (which is actually the physics
of the combination $w \rho^{2 |\nu|}$, since (\ref{bc1}) does not
depend on $w$ and $\rho$ independently).
In particular, in the $w,\rho$ parametrization here adopted one 
finds \cite{gacbak}
a simple formula that relates $w$ to $g$ once the reference 
scales $\rho$ and $\mu$ are fixed.

The realization that the class of deformations 
(\ref{bc1}) could be described using a contact interaction
also led \cite{gacbak} to a formulation of the anyon problem
in the powerful language of (non-relativistic) field 
theory, in which of course the contact interaction
is of the form $g \Phi^4$. (Here $\Phi$ is understood
as the bosonic field used in the ``bosonic gauge'' description
of anyons.) 
Interestingly, although the Calogero-Sutherland problem
is completely solved in the quantum-mechanical formulation,
we are yet unable to reformulate it as a \underline{local}
field theory. In practice, while we have indentified the Chern-Simons
field as the mediator of the exchange-statistical ``interaction'',
we (still ?) are unable to describe
exclusion-statistical interactions as mediated by a field.

Since it is by default set up as perturbative,
the Chern-Simons/anyon field-theoretical approach is directly connected
with the corresponding perturbative approach in the
quantum-mechanical formulation of anyons, and in fact these two perturbative
approaches are essentially the same thing, although 
one might be more convenient than the other from the point of view 
of computations, depending on the quantities of interest.
Of course, less direct is the correspondence between
these perturbative (contact-interaction-based) approaches
and the ``dual'' formulation in which the deformation
is enforced via the boundary condition (\ref{bc1}).
This ``duality'' has been tested by comparing the first few terms in 
the perturbative (contact-interaction based)
expansion of some quantities of interest to the 
corresponding approximation that can be obtained in the  
boundary-condition-deformation formulation, which in the 2-body
case can be solved exactly.
In light of the positive outcome of these tests
there is growing confidence in the ``duality'' between
contact-interaction formulation and boundary-condition formulation,
but, as also emphasized in \cite{gms}, one should probably keep open 
to the possibility that unexpected nonperturbative effects
might spoil the ``duality''.\footnote{It is probably worth emphasizing
that, while one should perhaps keep open to the possibility that 
nonperturbative effects might spoil the ``duality'',
one should \underline{not} misinterpret 
the above-mentioned ``delicate mathematics''
required to describe the special case $\nu = 0$ (within
the $w,\rho$ parametrization) as a signal of 
possible \underline{perturbative-level}
failures of the ``duality'' between contact-interaction formare of 
the formulation and boundary-condition formulation.
For example, in Ref.~\cite{gms}
a ``parametrization singularity'' that emerges in the $\nu \rightarrow 0$
limit has been \underline{erroneously} interpreted as a problem for 
this ``duality'', while that parametrization singularity
has no more physical content than (and is somewhat related to)
the well-known singularity that emerges
in trying to obtain the ${\cal H}$-eigenfunctions having
$\ln r$ behavior at small $r$ for $\nu \! = \! 0$ 
as a naive $\nu \rightarrow 0$ limit
of the ${\cal H}$-eigenfunctions with $r^{|\nu|}$ or $r^{-|\nu|}$
behavior at small $r$.}

One more aspect of the contact-interaction formulation
that deserves mention is the role of 
renormalization. Interestingly, this type of contact interactions
in non-relativistic quantum mechanics (or, equivalently, 
non-relativistic field theory) requires \cite{pap6p9,mto,gacbak,nuzero}
the full machinery of
renormalization that we are most accustomed to encounter in the context
of relativistic field theory.
The above-mentioned running coupling $g(\mu)$
emerges from the regularization/renormalization procedure.
In particular, one finds that the one-loop renormalization-group
equation satisfied by  $g(\mu)$ is
\begin{equation}
{d g \over d \ln \mu} = g^2 -\nu^2~.
\label{beta1}
\end{equation}
Of course the fixed points $g \! = \! \pm \nu$ 
correspond to the scale-invariant limits of the boundary
condition (\ref{bc1}) ($g \! = \! \nu \leftrightarrow w \! = \! 0$
whereas $g \! = \! - \nu \leftrightarrow w \! = \! \infty$).

The formalism and language of renormalization might be 
used also to reinterpret certain other aspects of this subject.
In particular, for the $\nu \! = \! 0$ case
one often finds in the literature statements to the
extent that the contact interactions discussed above can only 
be attractive. A similar statement is encountered in the 
$\nu \! \ne \! 0$ case, but there one allows for a small
repulsive contact interaction, small enough that the associated
repulsion does not overcome the ``centrifugal'' barrier
associated with the anyon anomalous spin.
These observations which arise in the mathematical setting of studies
of self-adjoint extensions are probably related to the ``triviality issues''
that one encounters in a renormalization group analysis.
For any given ultraviolet (short-distance) cut-off scale 
these models appear to be well defined for any value (however
positive or negative) of the contact coupling $g$.
As the cut-off is removed one finds that the repulsive models
are trivial, {\it  i.e.} they are not really
repulsive since their renormalized coupling vanishes in the limit
in which the cut-off is removed.
This is probably the renormalization-group counterpart
of the self-adjoint extension consistency arguments
finding that repulsive contact interactions are not allowed.
In physical applications the renormalization-group viewpoint
might be most relevant, since in most cases the models
here of interest only make physical sense with a finite cut-off.
For example, if anyons are seen as collective modes of a 
physical system they should also be well-defined only up to
a maximum energy-scale where the description
in terms of the fundamental degrees of freedom
becomes necessary.

I close this section by providing some evidence that 
another ``duality'' might be present in this framework,
a duality with some elements of the T-duality of strings
and perhaps even more closely related to the
Kramers-Wannier duality of 
the Ising model in two dimensions.
This evidence is found by examining
the structure of the two-particle relative partition function $Z_2$
that is associated to the two-particle relative 
Hamiltonian ${\cal H} + \omega^2 r^2$. 
(Of course, $Z_2$ is the only nontrivial
ingredient of the 2nd virial coefficient, so one could rephrase
what follows by emphasizing the implications 
for the 2nd virial coefficient.)

Since the eigenvalues of ${\cal H} + \omega^2 r^2$ are of the form
\begin{equation}
E_n = 2 \, (2n+1+ \Delta ) \, \omega ~,
\label{enern}
\end{equation}
where $\Delta$ depends on 
the boundary conditions ({\it i.e.} ${w_\rho}$) and $\nu$, $Z_2$ can 
be written as
\begin{equation}
Z_2 = \sum_{n=0}^{\infty} 
\exp[- 2 (2n+1+ \Delta )\beta\omega],
\end{equation}
Interestingly, all the dependence on $\nu$ and ${w_\rho}$ is
in the factor $\exp(- 2 \Delta \beta \omega)$, and it is therefore
plausible that
the same $Z_2$ be obtained for different combinations
of ${w_\rho}$, $\nu$, $\beta$, suggesting a duality
in which different boundary conditions
be connected by various coupling and temperature combinations. 
These dualities would characterize the full 
two-particle relative-motion sector of the Calogero-Sutherland problem,
but in the anyon problem $Z_2$ only takes into account the $L \! = \! 0$
projection of the two-particle relative motion, and for the full
result one would have to add
\begin{equation}
Z_2^{L\neq0} 
= \frac{\cosh[2 (1-\nu)\beta\omega]}{2\sinh^2[2 \beta\omega]} -
\frac{\exp(- 2 \nu\beta\omega)}{2\sinh[ 2 \beta\omega]} ~,
\end{equation}
which depends independently on $\nu$ and $\beta$.
The resulting properties of $Z_2 + Z_2^{L\neq0} $ have been 
discussed (however implicitly) in the recent Ref.~\cite{twovir},
and will be discussed in detail in Ref.~\cite{gacstef}.

\section{T-duality and boundary-non-critical strings}

Open strings 
provide another class of theories 
that admits interesting boundary-non-critical deformations.
As a first example in which to analyze the applicability to 
boundary-non-critical strings of T duality,
Ref.~\cite{gacnick} discussed the case of
a linear-dilaton boundary background.
Let me here review how that works in a 
flat 26-dimensional target spacetime.
Upon introduction of the
linear dilaton boundary background
in a non-critical string theory 
with central charge deficit $Q^2$ 
one is confronted by the action
\be
S= \frac{1}{4\pi {}}\int d^2 \sigma \partial X^i {\overline \partial} X^i
+ \frac{1}{4\pi {}}\int d^2 \sigma \partial Y^j {\overline \partial} Y^j 
- \int_{\partial \Sigma } Q {\hat k} \eta^i X^i 
\label{part}
\ee
where $\eta$ is an n-dimensional constant number-valued vector, and
(for later convenience)  I have divided the 26 fields into $n$ fields
of type $X$ ({\it i.e.} $i = 1 \dots n$) and $26-n$  fields of type $Y$
({\it i.e.}  $j = 1 \dots 26 -n$). I have also used
conventional notations ${\hat k}$ for the extrinsic curvature 
of the fiducial metric, 
$\Sigma$ for the world-sheet manifold, and $\partial \Sigma$
for the boundary of the world-sheet manifold.
The fields $X$ and $Y$ are assumed to satisfy Neumann boundary 
conditions: 
\be
  \partial_\sigma \hat X^i =  \partial _\sigma \hat Y^j =0 
\qquad {\rm on~}\partial \Sigma
\label{neuman}
\ee
where I use the notation $\partial_\sigma$ ($\partial_\tau$) 
for normal (tangential) derivatives on $\partial \Sigma$
and I also use the notation $\hat \Phi$ to emphasize that
a field $\Phi$ is being evaluated on  $\partial \Sigma$. 

The path-integral formulation of this $\sigma$-model is: 
\be
Z=\int DX \, DY \, \delta (\partial_\sigma \hat X^i)
\, \delta (\partial_\sigma \hat Y^j) \,
\exp \left[ -\int _{\Sigma} \frac{1}{4\pi{}} 
\left[ \partial X^i {\overline \partial} X^i +
\partial Y^j {\overline \partial} Y^j \right]
+ \int _{\partial \Sigma} {\hat k} Q \eta^i X^i \right]
~,
\label{opensigma}
\ee
where I indicated explicitly via boundary delta-functionals
that the $X^i$ and the $Y^j$ are Neumann fields.

As observed in Ref.~\cite{gacnick}
a functional T-duality transformation on the fields $X^i$
in the path integral (\ref{opensigma})
can be implemented by a straightforward
generalization of the formulation of T duality in the path-integral 
formalism for critical (Dirichlet or Neumann) open-string theories
(see {\it e.g.} Ref.~\cite{dornotto}).
In particular, also in the boundary-non-critical case
the T-duality transformation 
has as crucial element the introduction of a vectorial field variable
corresponding to the partial derivative of the fields $X^i$
that are being dualized:
\be
\yy^i_\alpha \equiv \partial_\alpha X^i ~.
\label{redef}
\ee
The fields $\yy$ are introduced in the path integral
via the identity:
\be
\int D\yy^i_\alpha \, \delta (\yy^i_\alpha-\partial_\alpha X)
\, \delta (\epsilon_{\alpha\beta}\partial_\alpha \yy^i_\beta)
=1 ~,
\label{constr}
\ee
which takes of course into account 
the `Bianchi identity':
\be
\epsilon_{\alpha\beta}\partial_\alpha \yy_\beta=0
\label{bianchi}
\ee
The path integral (\ref{opensigma}) is therefore rewritten as:
\bea
&~& Z= 
Z_Y \int DX^i \, D\yy^i_\alpha \, D\xd^i \, 
D\lambda^i_\alpha \, 
\delta (\partial_\sigma \hat X^i) \,
\exp \left[ \int_{\partial \Sigma} {\hat k} Q \eta^i X^i 
- i \int_{\partial \Sigma} \hat \lambda^i_\sigma \hat \yy^i_\sigma 
- i \int_{\partial \Sigma} \hat \lambda^i_\tau 
(\hat \yy^i_\tau -\partial_\tau \hat X^i) \right] 
\nn \\
&~&~~~~~~~~~~~~
\exp \left[
-\frac{1}{4\pi {}} \int_\Sigma  (\yy^i_\alpha)^2 
- i \int_\Sigma \xd^i (\epsilon_{\alpha\beta}\partial_\alpha \yy^i_\beta) 
- i \int_\Sigma \lambda^i_\alpha (\yy^i_\alpha-\partial_\alpha X) \right] 
\label{prepixtilde}
\eea
where (consistently with the notation already introduced for the normal
and tangential derivatives) I denoted the normal (tangential) components 
of world-sheet vectors with a lower index $\sigma$ ($\tau$). 
Summation is of course understood on all repeated indices apart 
from $\sigma$ and $\tau$ which are fixed labels for boundary fields.
I am also using the short-hand notation
\bea
&~& Z_Y \equiv \int DY \,
\delta (\partial_\sigma \hat Y^j) \,
\exp \left[
-\frac{1}{4\pi {}}\int_\Sigma \partial Y^j {\overline \partial} Y^j 
\right] 
\label{zy}
\eea
for the portion of the partition function that concerns
the $Y^j$ fields, which are ``spectators'' of the T-duality
transformation being performed on the $X^i$ degrees of freedom.
Moreover I adopt the convention $\epsilon_{\sigma\tau}=1$
and the following functional representation of a $\delta (\phi)$ constraint
\be
 \delta (\phi) = \int D\lambda e^{-i\int _{M} \phi \lambda}
\label{deltconstrfr}
\ee
with $M$ an appropriate manifold, indicating the range of definition of 
the arguments of the fields $\phi$. 
(In the cases here of interest $M=\Sigma$ or $\partial \Sigma$.)

The Lagrange multipliers fields 
$\xd^i$ and $\lambda^i_\alpha$ play a highly non-trivial
r\^ole in the T-duality
transformation; in particular, the fields $\xd^i$,
which implement the Bianchi identity (\ref{bianchi}), 
turn out to be
directly related to the fields that are T-dual to the fields $X^i$,
just as expected from the analysis reported in Ref.~\cite{dornotto}. 

As best seen by rearranging terms using integration by parts \cite{gacnick},
the functional integration over $X$ and $\yy$ can be done easily, and
one obtains (up to an irrelevant overall factor coming from the 
gaussian integration over $\yy$)
\bea 
Z= Z_Y \int D \xd \, D\lambda \, \delta (\partial_\alpha \lambda_\alpha)
\, \delta(\hat \lambda_\sigma) \, \delta (\hat \xd  - \hat \lambda_\tau) \,
\delta(\partial_\tau \hat\lambda_\tau + i Q {\hat k} \eta^i 
+ \hat \lambda_\sigma) 
\exp \left[-\pi {} \int_{\Sigma} (\lambda^i_\alpha 
- \epsilon_{\alpha\beta}\partial_\beta \xd^i)^2 \right] 
\label{pixtilde2}
\eea

The fields $X_D^i$ that are T-dual to the fields $X^i$
are easily identified as the ones satisfying the relation
\be
\frac{i}{2\pi{}} \epsilon_{\alpha \beta} \partial_\beta X^i_D \equiv 
\epsilon_{\alpha \beta} \partial_\beta \xd^i - \lambda^i_\alpha 
\label{xdualdef}
\ee
whose consistency follows from the 
constraint $\partial_\alpha \lambda^i_\alpha = 0 $ 
(see Eq.~(\ref{pixtilde2})).

Upon the change of variables $\xd^i \rightarrow X^i_D$,
and disposing of the then trivial functional integration over 
the $\lambda$ fields,
one can easily rewrite 
the partition function of (\ref{pixtilde2}) as
(up to another irrelevant overall factor)
\bea 
Z= Z_Y \int D X_D \, \delta(\partial_\tau \hat X^i_D + 2 \pi {}
Q {\hat k} \eta^i ) 
\exp \left[ - \int_{\Sigma} (\partial_\alpha X_D^i)^2 \right] 
\label{final1}
\eea
where I also used the fact that (\ref{xdualdef}), when combined
with the constraint $\hat \lambda^i_\sigma = 0$
(see Eq.~(\ref{pixtilde2})), implies 
$\partial_\tau \hat X^i_D = - i 2 \pi {} \partial_\tau \hat \xd^i$.

The T duality transformation implemented via the path-integral
manipulations that take from (\ref{opensigma}) to (\ref{final1})
evidently maps a Neumann open string with
boundary interactions corresponding to the
linear dilaton boundary background
present in (\ref{opensigma})
into a \underline{free} open string satisfying nonconformal
boundary conditions
\bea 
\partial_\tau \hat X^i_D = - 2 \pi {} Q {\hat k} \eta^i 
~.
\label{BCfinal1}
\eea
This boundary condition  
reduces to the Dirichlet boundary condition in the limit $Q \rightarrow 0$.
For every $Q \ne 0$ it encodes a ``conformal anomaly'' 
for the free T-dual theory that reflects the conformal anomaly 
of the corresponding boundary interactions of the original 
Neumann theory.

\section{Outlook}

Some of the topics here discussed 
might evolve very quickly in the coming years.
It would be important to establish in 
more general contexts (more general structures allowed on the boundary) 
the doings of T-duality transformations
in boundary-non-critical open-string theories.
It would also be interesting to find additional evidence
of the general interconnection between 
boundary non-critical quantum mechanics, T-type duality, and
(exchange or exclusion) exotic statistics
that appears to be suggested by the line of argument advocated in 
the present paper.

It is also tempting to consider the possibility that
some of the ideas of ``renormalization in quantum mechanics'' 
and  ``boundary-non-critical quantum mechanics''
might turn out to have  applications in the framework of
M(atrix) quantum mechanics~\cite{Mt}.
In particular, in the formulation adopted in Ref.~\cite{Mqm}
one ends up considering the effects of a $r^{-2}$ perturbation
in a 9-dimensional space and one is of course confronted
with infrared problems (and possibly ultraviolet problems if wave
functions with appropriate structure can be considered).
While the analysis reported in Ref.~\cite{Mqm} avoided
the regularization/renormalization machinery by advocating
some (rather formal) manipulations based on supersymmetry,
a rigorous treatment would require regularization/renormalization
({\it e.g.} of the type of the Hulthen-potential infrared regularization
discussed in Ref.~\cite{hult}).

\section*{Acknowledgements}
Several aspects of the analysis reported in this paper
have emerged in the context of collaborations with Dongsu Bak,
Stefan Mashkevich and Nick Mavromatos; their contribution
is gratefully acknowledged.
I also happily acknowledge useful discussions with
I.~Kogan, M.~Ortiz, many of the  
participants of the meeting {\it Quantum Aspects of 
Gauge Theories, Supersymmetry and Unification},
to which this volume is devoted,
and many of the participants of the meeting
{\it Common Trends in Condensed Matter and High Energy Physics}
(Chia, Italy, 1-7 September 1997), where this analysis
was first reported.
This work was supported in part by funds provided 
by the Foundation Blanceflor Boncompagni-Ludovisi, P.P.A.R.C.
and the Swiss National Science Foundation.


\end{document}